\documentclass[sigconf]{acmart}
\usepackage{subfig, float}
\usepackage{amsmath}
\usepackage{graphicx}
\usepackage{array}
\usepackage{balance}
\usepackage{colortbl}
\usepackage{makecell}
\usepackage{multirow}
\usepackage{bm}
\usepackage{wrapfig}
\usepackage[section]{algorithm}
\usepackage{algpseudocode}
\algnewcommand{\IfThenElse}[3]{
  \State \algorithmicif\ #1\ \algorithmicthen\ #2\ \algorithmicelse\ #3}
\algnewcommand{\IfThen}[2]{
  \State \algorithmicif\ #1\ \algorithmicthen\ #2}

\usepackage{tikz}
\usetikzlibrary{arrows}

\newtheorem{theorem}{Theorem}
\newtheorem{lemma}{Lemma}

\usepackage{xspace}
\newcommand{\mwp}{minimum winning price\xspace}

\newcommand{\Mwp}{Minimum winning price\xspace}

\newcommand{\hb}{\hat{b}}
\newcommand{\bx}{\boldsymbol{x}}
\newcommand{\bxi}{\boldsymbol{x}^{(i)}}
\newcommand{\balpha}{\boldsymbol{\alpha}}
\newcommand{\won}{\ell}

\usepackage[inline]{trackchanges}

\AtBeginDocument{%
  \providecommand\BibTeX{{%
    \normalfont B\kern-0.5em{\scshape i\kern-0.25em b}\kern-0.8em\TeX}}}

\setcopyright{acmcopyright}
\copyrightyear{2021}
\acmYear{2021}
\setcopyright{acmcopyright}
\acmConference[KDD '21] {Proceedings of the 27th ACM SIGKDD Conference on Knowledge Discovery and Data Mining}{August 14--18, 2021}{Virtual Event, Singapore.}
\acmBooktitle{Proceedings of the 27th ACM SIGKDD Conference on Knowledge Discovery and Data Mining (KDD '21), August 14--18, 2021, Virtual Event, Singapore}
\acmPrice{15.00}
\acmISBN{978-1-4503-8332-5/21/08}
\acmDOI{10.1145/3447548.3467167}

\begin{document}
\fancyhead{}

\title{An Efficient Deep Distribution Network for Bid Shading in First-Price Auctions}


\author{Tian Zhou, Hao He, Shengjun Pan, Niklas Karlsson, Bharatbhushan Shetty, Brendan Kitts, \\ Djordje Gligorijevic, San Gultekin, Tingyu Mao, Junwei Pan, Jianlong Zhang and Aaron Flores}

\renewcommand{\shortauthors}{\small T Zhou, H He, S Pan, N Karlsson, B Shetty, B Kitts, D Gligorijevic, J Pan, S Gultekin, T Mao, J Zhang and A Flores}

    \affiliation{ 
      \institution{Yahoo Research, Verizon Media}
      \city{Sunnyvale}
      \state{CA}
      \country{USA}
    }
    \email{{tian.zhou, hao.he, alanpan, niklas.karlsson, bharatbs, brendan.kitts, djordje, sgultekin, tingyu.mao}@verizonmedia.com,}
    \email{jonaspan@tencent.com, jianlong.zhang@gmail.com, aaron.flores@verizonmedia.com}

\begin{abstract}
Since 2019, most ad exchanges and sell-side platforms (SSPs), in the online  advertising industry, shifted from second to first price auctions.
Due to the fundamental difference 
between these auctions, demand-side platforms (DSPs) have had to update their bidding strategies to avoid bidding unnecessarily high and hence overpaying. \emph{Bid shading} was proposed to adjust the bid price intended for second-price auctions, in order to balance cost and winning probability in a first-price auction setup.
In this study, we introduce a novel deep distribution network for optimal bidding in both \emph{open (non-censored)} and \emph{closed (censored)} online first-price auctions.
Offline and online A/B testing results show that our algorithm outperforms  previous state-of-art algorithms in terms of both surplus and effective cost per action (\emph{eCPX}) metrics. Furthermore, the algorithm is optimized in run-time and has been deployed into VerizonMedia DSP as production algorithm, serving hundreds of billions of bid requests per day. Online A/B test shows that advertiser's ROI are improved by +2.4\%, +2.4\%, and +8.6\% for impression based (CPM), click based (CPC), and conversion based (CPA) campaigns respectively.
 
\end{abstract}



\maketitle
\section{Introduction}
\label{sec:introduction}
In online advertising, inventories that are not directly sold are primarily auctioned programmatically in real-time bidding (RTB). Before 2018, second-price was the dominant form of auction for RTB, where the winner only needs to pay the second highest bid price. However, starting in 2017, all the major exchanges/SSPs including AppNexus, Index Exchange, OpenX, Rubicon Project, and Pubmatic, with the exception of Google AdX, were rolling out or testing first-price auctions, where the winner must pay whatever bid it submitted, in varying degrees~\cite{sluis:bigchange}. Google transitioned to first-price auctions in 2019~\cite{google:fpa}. Several motivations were behind the transition away from second-price auctions. Firstly, first price auctions provided greater 
transparency and accountability for bidders, since the bidder was always charged exactly what they offered~\cite{chari1992us,sluis:guardiansuesrubicon,getintent:rtb1,rubicon:openletter}. Secondly, the unmodified second price auctions proved to be incompatible with  
the widespread and popular practice of Header Bidding~\cite{hearts2018:sold}. 

For demand-side platforms (DSPs), which are the bidders in the auctions, transitioning from second-price auctions to first-price auctions meant that bidding strategies would need to be dramatically adjusted. In second-price auctions, auction theory states that it is a dominant strategy for a bidder to bid \emph{truthfully}~\cite{easley2010}, namely it is the optimal strategy for a DSP to compute the value of the inventory being auctioned and submit this value as the bid price, regardless of the other bidders' behavior. However, for first-price auctions such a strategy would cause DSPs to overbid and thus lose money. This comes from the fundamental difference in the payment between second-price and first-price auctions. 

Unlike in second-price auctions, in first-price auctions a DSP must incorporate other bidders' behavior, more precisely its estimates of other bidders' bid prices, into its own bidding strategy. If the competing bidders' prices were known in advance, which is impossible in practice, the optimal bidding strategy would be to submit a bid price that is slightly higher than the highest competing bid price so as to win the auction with the lowest price possible. In reality, a DSP has to estimate the \mwp as best as it can, and lower its original bid price intended for second-price auction, i.e., \emph{shade} the truthful value of the inventory, accordingly. This process is known as \emph{bid shading}. Bid shading is relatively new to online advertising, but it has been used in auctions from other industries~\cite{chakravorti1995auctioning,capen:auctions,crespi2005multinomial,hortaccsu2018bid}.

The most important aspect of bid shading is a trade-off between the winning rate and (Return On Investment) ROI. The more the bid price is shaded, namely the lower the final bid price, the better the ROI if the bid is won. However, lower bid price also lead to lower probability of wining a bid. The buyer or bidder \emph{surplus}~\cite{hearts2018:sold,hortaccsu2018bid} is the shaded amount, namely
the difference between the bid price before shading and the final bid price if the bid is won. The surplus is 0 if the bid is lost. The quantitative objective of bid shading is thus to maximize the surplus, either directly or implicitly.

In the first price auction, an important piece of information is the \emph{minimum winning price}, which is the highest competing bid, inclusive of the floor price, if provided. However, it's up to an SSP to decide whether to provide the minimum winning price to the participating DSPs after the auction. \emph{Open (non-censored)} first price auctions refer to the auctions where feedback including \emph{\mwp} are shared to all participants regardless of auction outcome while in \emph{closed (censored)} auction, only the win or loss feedback is available. 

In this paper, we propose a deep distribution network to learn  the distribution of \emph{\mwp} for both censored and non-censored first price auctions, as well as for efficient search of the optimal bid price to maximize the profit in the real-time serving. The model has the flexibility of explicitly choosing the distributions of the \mwp and the structures of the network. Comprehensive experiments have been conducted and the model has been successfully deployed in one of the biggest DSPs in the world. We demonstrate the effectiveness of the algorithm by showing its performance lift compared to the existing models in offline and online setting.


To summarize, the main contributions of this study are the following:
\begin{itemize}
\item We propose a novel framework for bid shading that, to the best of our knowledge, is the first unified distribution-based bid shading method can be applied for both \emph{censored} and \emph{non-censored} first-price auctions.
\item We mathematically prove that an  efficient search algorithm can be used to find the optimal bid price that maximizes surplus given the distributions. This allows the model to be deployed online under strict latency constraints.
\item We implement and test the proposed algorithm in online and offline settings. The key metrics results show \textbf{+9.7\%} and {\textbf{+14.3\%}} lift for offline and online A/B tests respectively, compared to the existing models. Moreover, advertiser's ROI are improved by \textbf{+2.4\%}, \textbf{+2.4\%}, \textbf{+8.6\%} for impression based (CPM), click based (CPC), and conversion based (CPA) campaigns respectively.
\end{itemize}

The rest of the paper is organized as follows: In Section~\ref{sec:relatedwork}, we discuss related work in bid shading and online auctions. In Section \ref{sec:algorithms}, we describe the formulation of our bid shading algorithms, including details on how we train models to estimate the distributions of \emph{\mwp} and how to search for the optimal bid price efficiently on both \emph{censored} and \emph{non-censored} first-price auctions. In Section \ref{sec:offlineexperiments}, we  show the comprehensive experiments of different choices of distributions and network structures compared to the current state-of-art bid shading algorithms for both \emph{censored} and \emph{non-censored} first price auctions. 
In Section \ref{sec:onlineexperiments}, we briefly introduce the online deployment of deep distribution network in the serving system of our DSP and show the online ROI improvements compared to productions. Finally we conclude in Section  \ref{sec:conclusions}. 

\section{Related work}
\label{sec:relatedwork}
Bid optimization is one of the most fundamental problems in online advertising~\cite{zhou2008budget,yang2019bid,cai2017real,di2021unified,wu2018budget}. Recently, bid shading \cite{zulehner2009bidding,crespi2005multinomial,hortaccsu2018bid} attracts much attention since most ad exchanges and SSPs are shifting from second to first price auctions.
Bid Shading shares characteristics with the Seller's (Reserve) Price Optimization Problem \cite{kleinberg_selleropt, blum_selleropt, keskin_selleropt, segal_selleropt,simon_selleropt, kitts_selleropt}, where sellers have a good with a manufacturing cost, and their task is to set their price above this cost to maximize their profit. 
However, a variety of constraints exist in the bid shading buyer problem that are unique: 
(i) Buyers need to predict the bidding behavior of \emph{competing buyers} on each auction, leading to strategic considerations. (ii) Buyer Feedback is constrained in systematic ways, with censored and uncensored information provided. (iii) Buyers need to find a bid price for billions of auctions, each of which has combinatorial aspects; whereas sellers generally have a set of fixed inventory \cite{paul_personalizedpricing, wiki_personalizedpricing}. For these reasons, most authors talk about the Bidding problem as a Buyer specific activity, distinct from Seller reserve pricing, and we take the same approach in this paper.


There have been two general approaches for bid shading, depending on whether the \mwp is provided or censored. The first assumes that the \mwp is provided, and builds a machine learning algorithm to predict the \emph{optimal shading factor} - the ratio of the \mwp to the bid price before shading. For instance, Logistic Regression (\emph{LogReg}) and Factorization Machines (\emph{qFwFM})~\cite{bs:fwfm} have been used previously to predict the optimal shading factor, using an asymmetric loss function which penalizes losses. The drawback of these approaches is that they learn from the winnable bids only~\cite{bs:fwfm}, ignoring a large portion of available data. This paper proposes an algorithm for modeling distributions over the entire bidding landscape which allows for learning from both won and lost auctions.

The other general approach tries to estimate the distribution of the \mwp at segment level, and then finds the optimal bid price by maximizing the \emph{expected} surplus with respect to the estimated distribution. The \emph{NonLinear} algorithm in~\cite{niklas:shading} attempts to estimate the distribution using a non-linear approximation on a predefined segment. The main drawback of this method is that the distribution is estimated separately, and thus cross-segment information is not utilized. Furthermore, the segments must be explicitly defined and small segments must be manually grouped together. 
The recent WinRate model from~\cite{bs:winrate} takes a similar approach but approximates the distributions, implicitly with \emph{log-logistic}, for all segments simultaneously,
while its drawback is that it doesn't utilize the \mwp information when available.

Outside of the problem of bid shading, there has been some related research trying to characterize winning prices on auctions. For example, in~\cite{wu2015predicting, wu2018deep} the authors first estimate the winning price distribution, and then use it to do point-wise prediction of the winning price via a mixture model. The distribution implicitly assumes that the winning price follows logistic distributions, while a log-logistic distribution is used in~\cite{bs:winrate}. Additionally, there has been prior work estimating the bidding landscape in second-price auctions \cite{wu2015predicting, wu2018deep}, in which the \mwp feedback is only available while winning the auction. These approaches would need to be extended to work under first-price auctions, and the problem of surplus maximization.

Sell-Side Platforms are motivated to provide bid shading services to the bidders, especially during the transition period from second-price to first-price auctions. Such services include \emph{Bid Translation Service} from Google AdX~\cite{google:rtb}, \emph{Estimated Market Rate} from Rubicon Project~\cite{rubicon:EMR}, and \emph{Bid Price Optimization} system from AppNexus~\cite{appnexus:BPO}. However, 
these services are rather a transition tool in helping DSP's transition to first-price auctions, and many of them are being deprecated. For example, AdX deprecated its Bid Translation Service in May 2020~\cite{google:bts_deprecated}.

\section{Algorithms}
\label{sec:algorithms}
For the reader's convenience, we list some notations that will be used throughout the paper.\\[1ex]
\begin{tabular}{lp{0.85\columnwidth}}
\hline
$V$ & bid price before shading (true value of the ad opportunity) \\
$S$ & surplus \\
$s$ & expected surplus: $\mathbb{E}[S]$ \\
$b$ & bid price \\
$\hb$ & \mwp \\
$b^*$ & optimal bid price that maximizes the expected surplus $s$ \\
$D$ & distribution of the \mwp \\
$F(b)$ & Cumulative Distribution Function (CDF) of distribution $D$, i.e., winning probability at bid price $b$  \\
$f(b)$ & Probability Density Function (PDF) of distribution $D$: $F'(b)$ \\
$\bx$ & input feature vector that defines a segment \\
$\balpha$ & parameters of a distribution \\
\hline
\end{tabular} \\ [1ex] 
First of all, we define \emph{surplus} mathematically. Upon receiving a bid request for first-price auction, its value $V$ is  estimated based on its event rate, such as click-through rate (CTR) or conversion rate (CVR), and a campaign level price control signal that helps to pacing campaign budget smoothly across the flight. Let $b$ be the bid price to be submitted, and $\hb$ be the \mwp. $\mathbf{I}(b>\hb)$ equals 1 if $b>\hb$ and 0 otherwise, which indicates with bid price $b$ whether we win the auction. Then the \emph{surplus} is defined as
\begin{align}
    S(b; V,\hb) &= (V-b)\mathbf{I}(b>\hb)
    =\begin{cases}
    V-b, & \text{if }b > \hb, \\
    0, & \text{otherwise}.
    \end{cases}
\end{align}

Let $\bx=(x_1, x_2, \ldots, x_k)$ be the input feature vector derived from the current publisher and user attributes, such as top level domain, sub-domain, layout, day of week, etc. We calculate the optimal bid price $b^*$ given the current input feature vector in two steps:
\begin{description}
\item[Distribution Estimation] First we build a machine learning model to approximate the conditional distribution $D_{\hb\mid \bx}$ of the highest competing bid price $\hb$ by modeling its PDF or CDF.
\item[Surplus Maximization] Then we find the bid price $b=b^*$ that maximizes the expected surplus $\mathbb{E}_{\hb\mid \bx}[S]$:
\begin{align}
    b^* &= \mathop{\arg\max}_{b\in(0,V)}\,\mathbb{E}_{\hb\mid \bx}[S(b; V,\hb)] \nonumber\\
    & = \mathop{\arg\max}_{b\in(0,V)}\,\mathbb{E}_{\hb\mid \bx}\left[(V-b)\,\mathbf{I}(b>\hb)\right] \nonumber\\
    & =\mathop{\arg\max}_{b\in(0,V)}\,(V-b)\,\text{Pr}(\hb < b\, \mid x).\label{eq:max-surplus}
\end{align}
\end{description}

In the following sections, we describe in more details how the distribution estimation and surplus maximization are conducted.
\subsection{Inference of Distribution}
Conditioning on input feature vector $\bx$ of a bid request, to find the optimal bid price b* that maximizes the expected surplus, modelling of winning probability $\text{Pr}(\hb < b; \hb\mid \bx)$ for any intended bid price $b$ is the key. We assume that for given $\bx$, the \mwp follows a conditional distribution $D_{\hb\mid \bx}$. We further assume that each $\hb$ is drawn independently from a probability distribution $D_{\hb; \balpha(\bx)}$ that belongs to a family of known distributions, where $\balpha=(\alpha_1,\alpha_2,\ldots,\alpha_m)$ is its m-parameters vector. Note that the parameter vector $\balpha$ is a function of $\bx$, namely, bid samples with the same input feature vector follow the same distribution. We will discuss details about different distributions at greater length in Section \ref{sec:distributions}
\subsubsection{\Mwp is provided by SSPs}
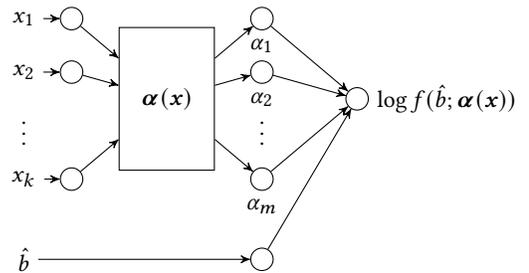
\begin{figure}[ht]
    \centering
    \begin{tikzpicture}[x=1em,y=2.25em,node distance=2em,>=stealth']
    \node (x1) at (0,0) {$x_1$};
    \node (x2) at (0,-1) {$x_2$};
    \node (dots) at (0,-2) {$\vdots$};
    \node (xk) at (0,-3) {$x_k$};
    \node (b) at (0,-4.5) {$\hb$};
    
    \node[circle,draw] (n1) [right of=x1] {} edge[<-] (x1);
    \node[circle,draw] (n2) [right of=x2] {} edge[<-] (x2);
    \node[circle,draw] (nk) [right of=xk] {} edge[<-] (xk);
    
    \node[rectangle,draw,minimum width=4em,minimum height=6em] (alpha)
        at (6,-1.5) {$\balpha(\bx)$}
        edge[<-] (n1) edge[<-] (n2) edge[<-] (nk);
    
    \node[circle,draw,label=below:$\alpha_1$] (a1) at (10,0) {}
        edge[<-] (alpha);
    \node[circle,draw,label=below:$\alpha_2$] (a2) at (10,-1) {}
        edge[<-] (alpha);
    \node (adots) at (10,-2) {$\vdots$};
    \node[circle,draw,label=below:$\alpha_m$] (am) at (10,-3) {}
        edge[<-] (alpha);
    \node[circle,draw] (nb) at (10,-4.5) {} edge[<-] (b);
    
    \node[circle,draw,label=right:$\log f(\hb;\balpha(\bx))$]
        at (14,-1.5) {}
        edge[<-] (a1) edge[<-] (a2) edge[<-] (am) edge[<-] (nb);
    \end{tikzpicture}
    \caption{Model structure for PDF estimation when the \mwp is provided by SSPs}
    \label{fig:alpha-pdf}
\end{figure}

In non-censored first-price auctions, the \mwp $\hb$ is provided by the SSP after the auction regardless of the outcome. We build a machine learning model to estimate the distribution of $D_{\hb\mid \bx}$. Let $\{(\bxi, \hb_i)\}_{i=1}^n$ be the training data set of bid samples, where $\bxi$ is the input vector, $\hb_i$ the \mwp provided by the SSP, and $n$ is the total number of training samples. Let $f(\hb; \balpha(\bxi))$ be the corresponding PDF. Then we estimate $\balpha$ using maximum likelihood estimation (MLE):
\begin{align}
\max_{\substack{\balpha\\ {\hb\mid \bx}\in\mathcal{D}}}&\quad\sum_{i=1}^n \log f\left(\hb_i; \balpha(\bxi)\right),
\end{align}
where $\mathcal{D}$ is a predetermined distribution family. The model structure for maximizing the log-likelihood is illustrated in Figure~\ref{fig:alpha-pdf}. Note that compared to $\bx$, $\balpha$ is of much lower dimension, namely $k\gg m$.


\subsubsection{\Mwp is censored by SSPs}
\begin{figure}[ht]
    \centering
    \begin{tikzpicture}[x=1em,y=2.25em,node distance=2em,>=stealth']
    \node (x1) at (0,0) {$x_1$};
    \node (x2) at (0,-1) {$x_2$};
    \node (dots) at (0,-2) {$\vdots$};
    \node (xk) at (0,-3) {$x_k$};
    \node (b) at (0,-4.5) {$b$};
    \node (won) at (0,-5) {$\won$};
    
    \node[circle,draw] (n1) [right of=x1] {} edge[<-] (x1);
    \node[circle,draw] (n2) [right of=x2] {} edge[<-] (x2);
    \node[circle,draw] (nk) [right of=xk] {} edge[<-] (xk);
    
    \node[rectangle,draw,minimum width=4em,minimum height=6em] (alpha)
        at (6,-1.5) {$\balpha(\bx)$}
        edge[<-] (n1) edge[<-] (n2) edge[<-] (nk);
    
    \node[circle,draw,label=below:$\alpha_1$] (a1) at (10,0) {}
        edge[<-] (alpha);
    \node[circle,draw,label=below:$\alpha_2$] (a2) at (10,-1) {}
        edge[<-] (alpha);
    \node (adots) at (10,-2) {$\vdots$};
    \node[circle,draw,label=below:$\alpha_m$] (am) at (10,-3) {}
        edge[<-] (alpha);
    \node[circle,draw] (nb) at (10,-4.5) {} edge[<-] (b);
    
    \node[circle,draw,label={[xshift=1.5em]$F(b; \balpha(\bx))$}]
        (F) at (14,-1.5) {}
        edge[<-] (a1) edge[<-] (a2) edge[<-] (am) edge[<-] (nb);
    \node[circle,draw] (nwon) at (14,-5) {} edge[<-] (won);
    \node[circle,draw,label=right:{$L\left(F(b; \balpha(\bx)),\won\right)$}]
        at (17.5,-1.5) {}
        edge[<-] (F) edge[<-] (nwon);
    \end{tikzpicture}
    \caption{Model structure for CDF estimation when the \mwp is censored by SSPs}
    \label{fig:alpha-cdf}
\end{figure}
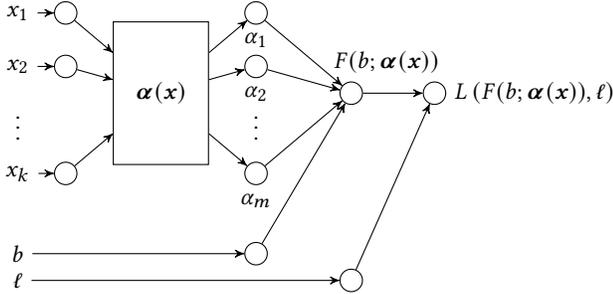

If the SSP does not provide the \mwp after the auction, the previous MLE approach will not work. In this case, we use an approach similar to that in~\cite{bs:winrate} to estimate the CDF of $D_{\hb\mid \bx}$, adding the flexibility of choosing the distribution family. Let $\{(\bxi, b_i, \won_i)\}_{i=1}^n$ be the training data set of bid samples, where $\bxi$ is the input feature vector, $b_i$ is the submitted bid price, and $\won_i\in\{0,1\}$ indicates if the bid was won. Let $F(b; \balpha(\bxi))=\text{Pr}(\hb_i < b; \balpha(\bxi))$ be the CDF of $D_{\hb\mid \bx}$. Then the likelihood of winning the bid with submitted price $b_i$ is $F(b_i; \balpha(\bxi))$. Since we know the result of the auction, we formulate it as a prediction problem. More precisely, we estimate $\balpha$ by minimizing the loss between the likelihood of winning and the actual result:
\begin{align}
\min_{\substack{\balpha\\ {\hb\mid \bx}\in\mathcal{D}}}
&\sum_{i=1}^n L\left(F( b_i; \balpha(\bxi)),\won_i\right),
\end{align}
where $\mathcal{D}$ is a predetermined distribution family, and $L$ is a loss function. In this paper, we use the well-known log loss:
\[
L\left(F(b; \balpha(\bx)),\won\right)
=\won \log F( b; \balpha(\bx))
+(1-\won)\log\left(1-F (b; \balpha(\bx))\right).
\]
The model structure for minimizing the log loss is illustrated in Figure~\ref{fig:alpha-cdf}. As can be seen, the win probability is first evaluated at bid price $b$ for given CDF and the loss is then calculated with the binary feedback $l$. 

In Section~\ref{sec:offlineexperiments} we will compare models built on different distribution families $\mathcal{D}$, as well as different structures of the block $\balpha(\bx)$.

\subsection{Surplus Maximization}
Regardless of whether the SSP provides the \mwp, let
\[s(b; V, \bx)\stackrel{\rm def}{=}(V-b)\text{Pr}(\hb < b;\, \hb\mid \bx)\]
be the expected surplus function with respect to $D_{\hb\mid \bx}$, the distribution of the \mwp for the current input feature vector $\bx$. Given $\bx$ and the bid price before shading $V$, the objective is to solve the following maximization problem:
\[
\max_{b\in(0,V)}\,s(b; V, \bx).
\]
If the underlying distribution $D_{\hb\mid \bx}$ results in a surplus function that has a unique local extremum (maximum or minimum), we can adopt the golden section search algorithm~\cite{kiefer:golden-section-search} shown in Algorithm \ref{alg:bisection} which converges in logarithmic time. It is a more numerically stable approach for most distributions, especially for gamma distribution and log-normal distribution, since no calculation of gradient is needed. The golden section search is more versatile and robust than other algorithms that require the calculation of gradients, which makes it more suitable for online implementation. 

\begin{algorithm}
\caption{Golden Section Search for Surplus Maximization}
\label{alg:bisection}
\begin{algorithmic}[1]
\Require \\
\begin{itemize}
    \item $V$: estimated value of the current ad opportunity
    \item $s(b)$: expected surplus function;
    \item $\epsilon>0$: minimum valid interval length
    \item $N$: maximum number of search steps
\end{itemize}
\Ensure $\beta > 0, V>0$, and
         $s(b)$ has exactly one local maximum and no local minimum in $(0,V)$.
\State $b_{\min},\, b_{\max} \leftarrow 0, V$
\State $gr \leftarrow (\sqrt{5}+1)/2$ 
\State $x_1\leftarrow b_{\max} - (b_{\max} - b_{\min})/gr$
\State $x_2\leftarrow b_{\min} + (b_{\max} - b_{\min})/gr$
\For{$i = 1, 2, \ldots, N$}
    \label{step:cut}
    \IfThenElse{$s(x_1) > s(x_2)$}{$b_{\max}\leftarrow x_2$}{$b_{\min}\leftarrow x_1$}
    \IfThen{$b_{\max} - b_{\min} < \epsilon$}{break} 
    \State $x_1\leftarrow b_{\max} - (b_{\max} - b_{\min})/gr$
    \State $x_2\leftarrow b_{\min} + (b_{\max} - b_{\min})/gr$
\EndFor
\Statex
\Return $(b_{\min} + b_{\max})/2$
\end{algorithmic}
\end{algorithm}

\subsection{Distribution Families}
\label{sec:distributions}
There are some constraints in choosing the distribution of minimum winning price to make economic sense. For example, its PDF should have a support between 0 and positive infinity. We mainly focus on  truncated-normal, exponential, gamma, and log-normal distribution families and show that the surplus functions for them have a unique local extrema, so that Algorithm~\ref{alg:bisection} is applicable to them all. We first introduce a few notations. Assuming that the \mwp $\hb$ follows a probability distribution $D$ with CDF $F(b)$ and PDF $f(b)$, then expected surplus and its first and second derivatives can be calculated as
\begin{align}
s(b) &= (V-b)\text{Pr}(\hb<b) =(V-b)F(b),
\label{eq:expsurplus} \\
s'(b) &= (V-b)f(b) - F(b), \label{eq:generalderv}\\
s''(b) &= (V-b)f'(b) - 2f(b). \label{eq:generalderv2}
\end{align}
For simplicity, we don't explicitly write the above as functions of input feature vector $\bx$ and bid price before shading $V$ here.

We first show that for all truncated-normal, exponential, gamma, and log-normal distributions, $s''(b)$ has at most one root in $(0,V)$. Let
\begin{align*}
f_{tn}(b) &= \frac{\exp\left(-\frac{(b-\mu)^2}{2\sigma^2}\right)}{\sqrt{2\pi}\sigma\left[\Phi\left(\frac{B-\mu}{\sigma}\right)-\Phi\left(\frac{A-\mu}{\sigma}\right)\right]}, \text{ where }\sigma>0 \text{ and}\\
&\quad -\infty\le A<B\le\infty,\tag{truncated-normal}\\
f_e(b) &= \lambda e^{-\lambda b}, \text{ where }\lambda > 0, \tag{exponential}\\
f_g(b) &= \frac{\beta^\alpha}{\Gamma(\alpha)}b^{\alpha-1}e^{-\beta b}, \text{ where }\alpha>0,\beta>0, \text{ and} \tag{gamma}\\
f_{ln}(b) &=\frac{1}{\sqrt{2\pi}\sigma b}\exp\left(-\frac{(\ln b-\mu)^2}{2\sigma^2}\right), \text{ where }\sigma>0. \tag{log-normal}
\end{align*}
Throughout this paper, we use the truncated-normal distribution with $A=0$ and $B=\infty$. Thus, for all these four distributions, the support of $b$ are $[0, \infty]$.

\begin{lemma}\label{lem:spp-one-root}
For any truncated-normal, exponential, gamma, or log-normal distribution of the \mwp, $s''(b)$ has at most one root in $(0,V)$. Furthermore, $s''(V)<0$.
\end{lemma}
\begin{proof}
Using Equation~\eqref{eq:generalderv2}, we can calculate $s''(b)$ for each of the listed distributions.
\begin{itemize}
\item For truncated-normal distribution we can show that
\begin{align*}
s''(b) &= f_{tn}(b)\left[(V-b)(\mu-b)\sigma^{-2}-2\right]. 
\end{align*}
Let $g(b)=(V-b)(\mu-b)\sigma^{-2}-2$.
Note that $g(b)$ is quadratic and hence convex, and $g(V)=-2<0$. Thus $g(b)$ has at most one root in $(0,V)$, and so does $s''(b)$.

\item For gamma distribution we can show that
\begin{align*}
s''(b) &= \frac{f_g(b)}{b}\left[(\alpha-1-\beta b)(V-b) - 2b\right].
\end{align*}
Let $g(b)=(\alpha-1-\beta b)(V-b) - 2b$.
Again, $g(b)$ is quadratic and hence convex, and $g(V)=-2V<0$. Thus $g(b)$ has at most one root in $(0,V)$, and so does $s''(b)$.

Since exponential distribution is a special case of gamma with $\alpha=1$ and $\beta=\lambda$, the same proof holds for exponential distribution.

\item For log-normal distribution we can show that
\begin{align*}
s''(b) = \frac{f_{ln}(b)}{b\sigma^2}\left[(\mu-\sigma^2-\ln b)(V-b)-2\sigma^2 b\right].
\end{align*}
Let $g(b)=(\mu-\sigma^2-\ln b)(V-b)-2\sigma^2 b$.
It's easy to verify that $g''(V)=(b+V)/b^2>0$, so $g(b)$ is convex. Since $g(V)=-2\sigma^2V<0$, $g(b)$ has at most one root in $(0,V)$, and so does $s''(b)$.
\end{itemize}
In summary, for all the listed distributions the corresponding $s''(b)$ has at most one root in $(0,V)$, and $s''(V)<0$.
\end{proof}
Finally we show that for all the listed distributions the corresponding surplus function has one global maximum.
\begin{theorem}\label{thm:s-one-extremum}
For any truncated-normal, exponential, gamma, or log-normal distribution of the \mwp, the surplus function $s(b)$ has one global maximum and no local minimum in $(0,V)$.
\end{theorem}
\begin{proof}
From Lemma~\ref{lem:spp-one-root}, for any truncated-normal, exponential, gamma, or log-normal distribution, $s''(b)$ has at most one root in $(0, V)$ and $s''(V)<0$. Then only one of the following two cases could happen:
\begin{itemize}
    \item $s''(b)<0$ for all $b\in(0,V)$, in which case $s(b)$ is concave and hence has at most one local maximum and no local minimum.
    \item $s''(b)$ has a unique root $b_0\in(0,V)$ such that $s''(b)>0$ for all $b\in(0,b_0)$ and $s''(b)<0$ for all $b\in(b_0,V)$, in which case $s(b)$ is convex in $(0,b_0)$ but concave in $(b_0,V)$. Note that $s(0)=s(V)=0$, then $s(b)$ must have at most one local maximum in $(b_0, V)$ and no local minimum.
\end{itemize}
In both cases, $s(b)$ has at most one local maximum and no local minimum. Further, since $s(0)=S(V)=0$ and $s(b)>0$ for all $b\in(0,V)$, $s(b)$ must have one global maximum.
\end{proof}

\section{Offline Experiments}
\label{sec:offlineexperiments}
In this section, we present comprehensive offline experiments on our DSP private bidding dataset. The following questions would be answered in the following sub-sections:
\begin{itemize}
\item \textbf{Q1:} Does lower log-loss or better distribution fit results in a higher surplus?
\item \textbf{Q2:} How much is the surplus lift when \mwp is available in training the deep distribution network compared with when \mwp is not available?
\item \textbf{Q3:} Will performance be improved by using more powerful network structures (deepFM etc.) that capture high-order feature interactions compared to logistic regression or Factorization Machines (FM)?
\end{itemize}
\subsection{Dataset}
The dataset we use for offline experiments is VerizonMedia DSP private bidding dataset on Adx exchanges. We extracted 12 fields through feature engineering, including exchange id, top level domain of the ad opportunity, sub domain, layout of the ad, position of the ad, device type, name of app, publisher id of the ad request, country, user local hour of the day, user local day of the week, if the user is new. There are billions of records, with millions of active features and minimum winning price available. We use 7 day's data to train the model and use 1 day's data to test.  Notice that we train 
two types of models separately to simulate censored and non-censored first-price auctions. For the pdf estimation model which introduced in Fig. \ref{fig:alpha-pdf}, minimum winning price are used as labels, while for the cdf estimation model in Fig. \ref{fig:alpha-cdf}, the binary win or lose information  (1/0) is used as label. We used the bid requests in the past 7 days for training to mitigate impact of the day of week pattern. Log-loss and surplus are the main two metrics we use in the offline experiment. 

\subsection{\Mwp provided by SSPs}
The production bid shading algorithms for non-censored first-price auctions is Factorization Machine based point estimation algorithm. It uses Field-weighted Factorization Machine (FwFM) as the model structure and learns the optimal shading factor for each bid request with an asymmetric loss function (penalize more when losing the bid) ~\cite{bs:fwfm}. To have a fair comparison with the baseline, we uses same model structure: FwFM as the model structure in Figure~\ref{fig:alpha-pdf}, and conduct experiments of different distributions we introduced in Section \ref{sec:algorithms}.3. For surplus metric, we only show the percentage of lift compared to the production model for privacy reason. \\
Table \ref{tbl:pdf_training} summarizes the results and we do observe the correlation between log-loss and surplus, in the sense that lower log-loss results in higher surplus. Among all the distributions, log-normal has the best performance with 9.7\% surplus lift compared to the current production algorithm possibly due to its capability to better model the long-tail distribution of the \mwp. 

\begin{table}[ht]
\centering
\begin{tabular}{lcc}
\toprule
Model & Log loss & Surplus Lift Percentage   \\
\midrule
truncated-normal & 0.87 & 1.45\%  \\
exponential     & 0.68 & 5.12\% \\
gamma           & 0.58 & 6.85\% \\
log-normal       & 0.56 & 9.65\% \\
\bottomrule
\end{tabular}
\caption{Performance of different distributions compared to production FwFM algorithm where highest competing bid price is not censored by the SSP}
\label{tbl:pdf_training}
\vspace{-5pt}
\end{table}

\subsection{\Mwp not provided by SSPs}
To simulate the censored first-price auctions where \mwp feedback is not available. We train the deep distribution network on the same dataset while not using \mwp information during training. The loss function is defined in Section 3.1.2. The production algorithm on censored SSPs is win-rate distribution algorithm, in which winning probability function is estimated as a sigmoid function ~\cite{bs:winrate}.
\begin{table}[ht]
\centering
\begin{tabular}{lcc}
\toprule
Model & Log loss & Surplus Lift Percentage  \\
\midrule
truncated-normal & 1.14 & 0.10\% \\
exponential     & 0.89 & 3.29\% \\
gamma           & 0.72 & 3.58\% \\
log-normal       & 0.68 & 5.32\% \\
\bottomrule
\end{tabular}
\caption{Performance of different distributions compared to production win-rate distribution algorithm}
\label{tbl:cdf_training}
\vspace{-5pt}
\end{table}

The results are presented in Table \ref{tbl:cdf_training}, and we observe the same correlation between log-loss and surplus again, which answers \textbf{Q1}. Notice that the log-loss is not applicable to the production algorithm since its definition of loss function is different from our model. Additionally, in comparison with non-censored training results in Table \ref{tbl:pdf_training}, there is a slightly drop in surplus performance lift for all distributions. This shows the importance of \mwp feedback in deep distribution network training which answers \textbf{Q2}. 

As can be seen from the offline results corresponding to the cases with \mwp in Table \ref{tbl:pdf_training} and the cases without \mwp in Table \ref{tbl:cdf_training}, log-normal distribution results in the lowest log-loss and the highest surplus. It indicates that among all the above distributions that we considered so far, log-normal fits the \mwp better. Therefore, we will choose log-normal as the output layer distribution in the online A/B test in Section \ref{sec:onlineexperiments}. 

\subsection{Network structure comparison}
\label{subsec:link}
In this subsection, we present the results on comprehensive experiments on applying well-known click-through rate prediction models like FM, FwFM, DeepFM, Wide \& Deep~\cite{ctr:FM, ctr:FwFM, ctr:deepFM, ctr:widedeep, pan2019predicting,ctr:deeplight}, to deep distribution network, as the deep distribution network structure, in learning the distribution of \mwp~. The results are shown in Table \ref{tbl:link_functions} which answers \textbf{Q3}. It can be seen that deep models, like deepFM, wide \& deep, which capture high-order feature interaction are able to learn a better distribution with relatively low log-loss, and thus have a better surplus compared to linear model LR, and shallow models FM, FwFM.

\begin{table}[ht]
\centering
\begin{tabular}{lcc}
\toprule
network structure & Log loss & Surplus Lift Percentage    \\
\midrule
LogReg & 0.718 & / \\
FM & 0.569 & 4.52\% \\
FwFM   & 0.558 & 4.70\% \\
wide \& deep   & 0.522 & 6.36\% \\
deepFM  & 0.521 & 7.10\% \\
\bottomrule
\end{tabular}
\caption{Performance of different network structures with lognormal as pre-defined distribution}
\label{tbl:link_functions}
\vspace{-5pt}
\end{table}

Based on the offline experiments results and online latency constraint, we eventually decide to use FwFM as the network structure and log-normal distribution for online experiments. However, we show the potential of more complex deep network structure, if the online latency requirements can be satisfied.

\section{Online Experiments}
\label{sec:onlineexperiments}
In this section, we will briefly introduce the real-time bid shading serving module we implemented, which serves billions of bid requests per day in one of the world's largest DSP. The online experimental results, including Return on Investment (ROI) for advertisers, are also presented, which further prove that deep distribution network outperforms other existing bid shading algorithms in literature.
\subsection{Bidding System Overview}
\begin{figure}[h]
	\centering
	\includegraphics[width=\columnwidth]{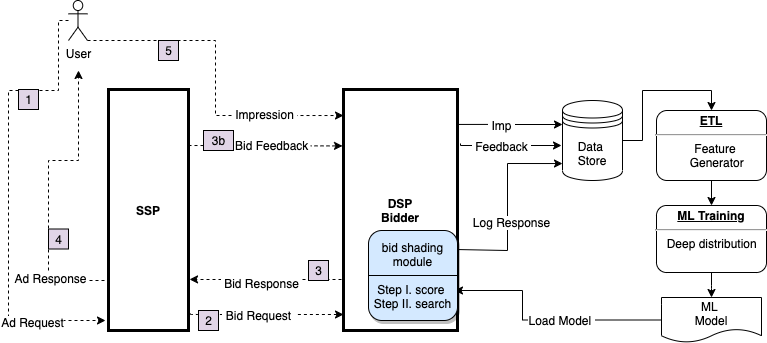}
	\caption{An Overview of VZDSP Bid Shading System}
	\label{fig:system_overview}
\end{figure}

Our DSP is a single platform that brings programmatic, premium, and its native marketplace inventory, formats, targeting and measurement together. We provide an overview of the bid shading aspects of the system, as illustrated in Figure~\ref{fig:system_overview}.

When a user visits a web page with an ad opportunity, an ad request is sent to the SSP responsible for selling it. The SSP then packages available user and page information into bid requests and send them to multiple DSPs for auction. Upon receiving a bid request, within hundreds of milliseconds our bidding system goes through various stages such as fetching user profiles, ads targeting, making click or conversion predictions, etc., and eventually estimates the \emph{true value} of the underlying ad opportunity for a selected candidate ad, or multiple candidates. 

For second-price auctions, this \emph{true value} would be the bid price. For first-price auctions, we have an additional \emph{bid shading module}, as shown on the right-hand side of the flow chart. In this bid shading module, historical bid information, including impressions and bid feedback from SSPs, are used to generate features and train bid shading models. The model is periodically updated and loaded into the real-time bidding system, and used to shade the \emph{true value} to produce the final bid price, which, as part of the bid response, is sent back to the SSP.

The SSP collects bid responses from all participating DSPs, determines the winner, and sends an ad response to the web page. If our DSP wins the auction, an impression with the selected ad would be shown on the user's web page, and relevant user information would be sent back to us. The SSP also sends bid feedback to its bidders, but it may or may not include the \mwp.

\subsection{Online Evaluation Metrics}
The most important key performance metrics for campaign delivery that the DSP system seeks to optimizes towards are CPM (Cost per thousand impressions), CPC (Cost per click) and CPA (Cost per action).  
For our online A/B test experiments of the bid shading algorithms described in this paper, we mainly focus on campaigns with one of those 3 optimization goal types,  which cover more than 80\% of total DSP revenue. For a given campaign spend budget, the DSP optimization system seeks to minimize the campaign's CPM, CPC or CPA as indicated by its optimization goal. For convenience, we use the term effective cost per event (eCPX) to denote either CPM, CPC or CPA. \\
Since optimization goals are defined at the campaign level, there is a need to define metrics to measure impact of algorithm improvements (in our case bid shading) across multiple campaigns. Simple aggregations of events (e.g. actions) and cost across campaigns to define a simple aggregated eCPX metric are not good metrics, since a few campaigns can dominate such metrics if it turns out they have orders of magnitude more events than the rest of the campaigns, even though their cost (campaign spend) is as high as any other campaign's spend. Moreover, it's not fair to compare two algorithms when their spends are different since the one with more spend tends to have higher eCPX. Thus, we are going to introduce two novel online DSP business-related metrics: campaign level eCPX statistics, and Bidder Performance Index (BPI), which can be used to measure improvements in campaign performance and ROI for our advertisers.

\subsubsection{\textbf{Campaign Level eCPX statistics}}
\begin{figure}
\centering
\subfloat{\includegraphics[trim=0cm 0cm 0cm 0cm,clip, width=.8\columnwidth]{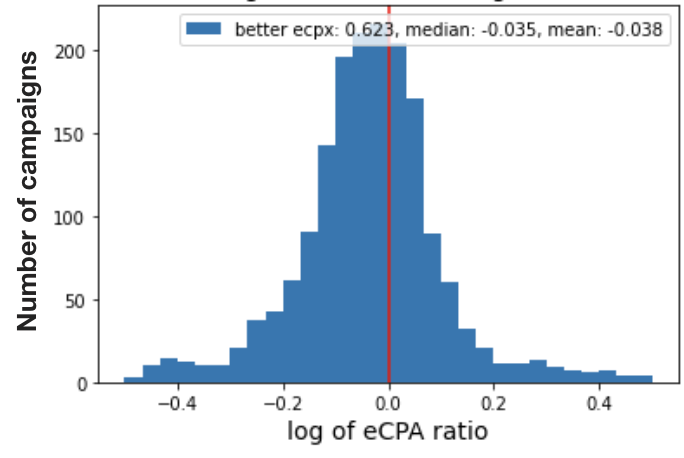}}
\caption{Histogram of log eCPX ratio: CPA campaign as an example}
\label{fig:ecpx_campaign_histogram}
\end{figure}

\begin{figure}
\centering
\subfloat{\includegraphics[trim=0cm 0cm 0cm 0cm,clip, width=.8\columnwidth]{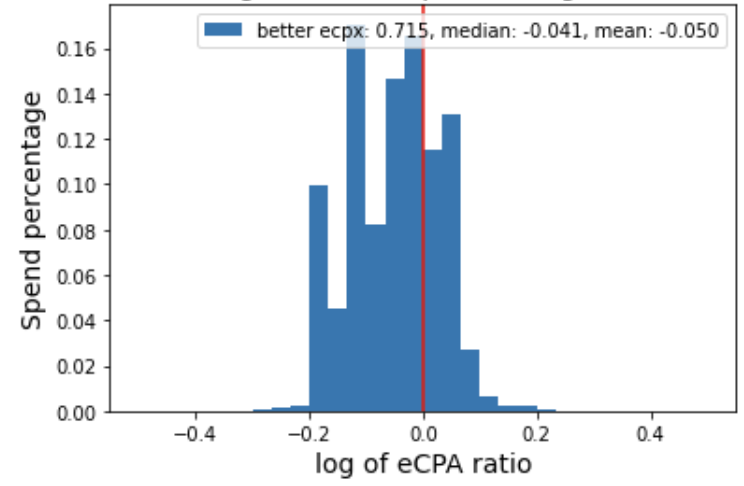}}
\caption{Histogram of log eCPX ratio weighted by campaign spend: CPA campaign as an example}
\label{fig:ecpx_histogram}
\end{figure}

A natural way to put together eCPX metrics across campaigns is to generate statistics on improvements measured at the campaign level, i.e. based on the eCPX of control and test buckets for each campaign. To avoid the asymmetry implied by the usual percentage difference in eCPX between control and test, we define the log of eCPX ratio for each campaign as the following, 
$$    r_{\text{ecpx}} = \log\left(\frac{\text{eCPX}_{\text{Test}}}{\text{eCPX}_{\text{Control}}}\right). $$
For a campaign that has similar spend in both control and test bucket, if $r_{\text{ecpx}} < 0$, then the test algorithm outperforms the production model. This provides a metric of improvement (or deterioration) in eCPX that is symmetric in the sense that the absolute value of the metric remains unchanged if control and test roles are interchanged.
A histogram depicting the distribution of $r_{\text{ecpx}}$ is shown in Fig \ref{fig:ecpx_campaign_histogram} for CPA goal type campaigns. Based on the $r_{\text{ecpx}}$ distribution, we can measure that 71.5$\%$ of campaigns have better eCPX, and the median and mean of $r_{\text{ecpx}}$ are, respectively, $-0.035$ and $-0.038$. Since campaigns can have very different budget and spend, the weighted histogram of $r_{\text{ecpx}}$ is shown in Fig \ref{fig:ecpx_histogram} where the weights are the spend of each campaign.

\subsubsection{\textbf{Bidder Performance Index (BPI)}}
Minimizing eCPX for a given spend amount is equivalent to maximizing value or Return to advertisers under the same amount of spend (Cost), where Return to advertisers is the monetary value of the events (impressions, clicks or actions) driven by the campaign. This motivates the so-called BPI metric, defined as: 
$$\frac{(\text{Return}_{\text{test}} - \text{Cost}_{\text{test}}) - (\text{Return}_{\text{control}} - \text{Cost}_{\text{control}})}{\text{Cost}_{\text{control}}}$$
\noindent where $\text{Return}$ and $\text{Cost}$ can be aggregated across campaigns. The numerator shows the extra $\text{Return}$ the new algorithm brings to the advertiser compensated by the extra cost it may also incur. Notice that it can be negative if the test algorithm is no better than the control one. The BPI metric is an aggregation-based type of metric that is less prone to be dominated by campaigns with very large number of events as explained before.


\subsection{Online A/B test results}
After rolling out the deep distribution network algorithm with FwFM network structure and log-normal distribution on Adx, one of the largest SSPs, we were able to monitor its online performance by maintaining a percentage of traffic that was randomly allocated to each algorithm. There are thousands of campaigns running everyday in our DSP, the summary of these campaigns under our A/B test environment during 3 consecutive days is shown in Table \ref{tbl:abtest}. 

It can be seen that for all goal types, the median of $r_{ecpx}$ are negative values, showing a overall better performance of our proposed algorithm. The weighted median, where the weights are the spend of each campaign, are also negative across all goal types. 
A majority of campaigns have a better eCPX performance on test bucket as can be seen from the third row of better campaign percentages. And These campaigns take up to 70\% to 80\% of the total spend. 
In all, from the online A/B test results, it can be shown that our proposed bid shading algorithm generates a significantly better performance than the current model in production.

\begin{table}[ht]
\resizebox{0.43\textwidth}{!}{%
\begin{tabular}{lccc}
\toprule
\multirow{2}{*}{Metric} & \multicolumn{3}{c}{campaign goal type} \\
       \cline{2-4}
       & CPA & CPC & CPM  \\
\midrule
median of $r_{\text{ecpx}}$                 & -3.5\% & -0.9\% &  -6.2\%  \\
median of spend weighted $r_{\text{ecpx}}$   & -4.1\% & -5.4\% &  -5.7\%  \\
better campaigns              & 62.3\% & 54.5\%   &  81.6\%  \\
better spend                  & 71.5\% & 80.6\%   &  71.8\%  \\
BPI                           & +8.57\%& +2.35\%  &  +2.35\% \\
\bottomrule
\end{tabular}}%
\caption{Online A/B test performance by goal types}
\label{tbl:abtest}
\vspace{-5pt}
\end{table}

\section{Conclusions}
\label{sec:conclusions}
In this work, we propose the deep distribution network for bid shading, which can be applied for both censored and non-censored first-price auctions. The parametric conditional distribution of \mwp is learnt through a FwFM network based on selected features of the bid request. For several well-known distributions, we proved that the resulting surplus function has a unique local maximum. Based on such property, an efficient golden-section search algorithm was applied at the real-time in finding the optimal bid price that maximizes expected surplus. In offline experiments, the proposed model out-performed existing state of the art bid shading algorithm on both censored and non-censored scenarios with respectively 9.7\% and 5.3\% surplus lift.  
Online A/B test showed that the proposed algorithm increases surplus by +14.3\% and brings +2.4\%, +2.4\%, +8.6\% ROI lift for impression based (CPM), click based (CPC), and conversion based (CPA) campaigns respectively.
The deep distribution network has been successfully deployed in one of the biggest DSPs in the world, serving billions of bid requests every day. Another major advantage of our framework is that, the overall structure with a deep neural network and a parametric distribution output layer, can be easily generalized. In the future, we will continue exploring more powerful and run-time efficient neural network structures, combining with other single or multi mode distributions for further improved performance.






\balance 
\bibliographystyle{ACM-Reference-Format}
\bibliography{references}


\end{document}